\documentstyle[aps,preprint,eqsecnum]{revtex}

\makeatletter
                \def\references{%
                \ifpreprintsty
                \vspace{2cm}
                \hbox to\hsize{\hss\large \refname\hss}%
                \else
                \vskip24pt
                \hrule width\hsize\relax
                \vskip 1.6cm
                \fi
                \list{\@biblabel{\arabic{enumiv}}}%
                {\labelwidth\WidestRefLabelThusFar  \labelsep4pt %
                \leftmargin\labelwidth %
                \advance\leftmargin\labelsep %
                \ifdim\baselinestretch pt>1 pt %
                \parsep  4pt\relax %
                \else %
                \parsep  0pt\relax %
                \fi
                \itemsep\parsep %
                \usecounter{enumiv}%
                \let\p@enumiv\@empty
                \def\theenumiv{\arabic{enumiv}}%
                }%
                \let\newblock\relax %
                \sloppy\clubpenalty4000\widowpenalty4000
                \sfcode`\.=1000\relax
                \ifpreprintsty\else\small\fi
                }
\makeatother

\begin{document}
\preprint{CERN-TH/97-4}
\title{	Calculation of higher-twist evolution kernels for\\
	polarized deep inelastic scattering}
\author{
        D. M\"uller
        }
\address{
       	CERN, Geneva, Switzerland\footnote{
	Permanent address: Institut f\"ur Theoretische Physik, 
	Universit\"at Leipzig, 04109 Leipzig, Germany}
        }
\maketitle

\begin{abstract} 
Based on the non-local light-ray operator technique, we develop an algorithm
for the computational calculation of evolution kernels for higher-twist
operators in leading order of the perturbation theory. We compute the
evolution kernel for the twist-3 operators in the flavour singlet channel.
Our result confirms the local anomalous dimensions computed by Bukhvostov,
Kuraev, and Lipatov as well as the non-local evolution kernels of Balitzky
and Braun. 
\end{abstract}

\section{Introduction}
\label{introduction}

Recently, in deep inelastic scattering (DIS) the first moments of the
transverse polarized structure function $g_2(x_{Bj},Q^2)$ have been measured
\cite{SMC94E143}. Among the leading twist-2 part $g_2(x_{Bj},Q^2)$
\hfill contains \hfill a non-power \hfill suppressed \hfill twist-3\hfill
part, \hfill which \hfill it \hfill is \hfill intended \hfill to \hfill
measure\\
\vfill
\noindent
CERN-TH/97-4 \\
\noindent
January 1997
\newpage
\noindent
at HERMES and the SMC. Although at present the statistic is too low to extract
the evolution of the twist-2 part of $g_2(x_{Bj},Q^2)$ with respect to the
momentum transfer $Q^2$, it may be important for future high-precision
measurements to know the theoretical prediction also for the twist-3 part.
This higher-twist contribution is in fact expressed in terms of three
particle operators sandwiched between the polarized nucleon state;
therefore, it possesses no simple parton interpretation. Furthermore, also
for (higher-twist) operators with fixed spin there remains a mixing problem
due to the renormalization. Consequently, the evolution equation is no
longer of the type of a Dokshitzer-Gribov-Lipatov-Altarelli-Parisi (DGLAP)
equation.

Introducing so-called qausi-partonic operators, Bukhvostov, Kuraev and
Lipatov derived evolution equations for higher-twist correlation functions
and computed the twist-3 evolution kernels and the anomalous dimensions for
both the non-singlet and the singlet channels
\cite{BukKurLip83bBukKurLip84a,BukKurLip84b}. Their non-singlet result was
confirmed in the axial gauge as well as in the covariant gauge by different
authors using quite different techniques. The results in
\cite{Rat86,BalBra89} coincide with the original result; however, in
\cite{JiCho90} the transposed local anomalous dimension matrix was obtained.
Recently, two more independent calculations
\cite{KodYasTanUem96,GeyMueRob96a} support the Bukhvostov, Kuraev and
Lipatov result.

It should be mentioned that special emphasis was given in
\cite{KodYasTanUem96} (see also \cite{KodYasUem95}) to the equation of
motion (EOM) operators. As is expected from the renormalization properties
of gauge-invariant operators \cite{JogLee76}, the result obtained in the
off-shell scheme coincides with the previous ones in which the mixing with
the EOM operators was neglected from the beginning of the calculation. In
the minimal subtraction scheme this result is obvious at leading order.
However, beyond the leading order it is not excluded that a non-proper
regularization of infrared singularities spoils the general renormalization
principles.

Here we apply an appropriate technique, which is based on the non-local
operator product expansion introduced by Anikin and Zavialov \cite{AniZav78}
and the renormalization group equation for light-ray operators
\cite{BorGeyHorRob85}. For several years Geyer, Robaschik and others applied
this approach to the calculation of leading twist non-forward evolution
equations \cite{BraGeyRob87} relevant for deeply inelastic Compton
scattering and, more recently, for the forward twist-3 non-singlet channel
\cite{GeyMueRob96a}. A similar approach was independently developed by
Balitzky and Braun \cite{BalBra89} and among other studies, applied to the
one-loop calculation of the non-forward twist-3 singlet channel.

Although the singlet result \cite{BalBra89} coincides with the local
anomalous dimension matrix given in \cite{BukKurLip84b} (up to obvious
misprints in both papers) it is desired to have a third independent
calculation. On the other hand it is also theoretically interesting to study
the evolution of power-suppressed contributions. Calculating the first few
moments of twist-4 operators, a first step in this direction was done in
\cite{BukFro87}. It should be stressed that also one-loop order calculations
are cumbersome and will be more difficult for increasing twist. Thus, it
would be desired to have an algorithm that allows the use of computer power.
Such an algorithm will be given in this paper for light-ray operators.

\section{Higher-twist contributions to transverse polarized DIS}
\label{Twist-3}


Applying the non-local operator product expansion
\cite{BorGeyHorRob85,GeyDitHorMueRob94,BalBra89} the structure functions for
DIS are factorized in coefficient functions and forward matrix elements of
light-ray operators. The well-known result for the longitudinal structure
function $g_1$ reads in leading order
\begin{eqnarray}
\label{LOg1}
g_1(x_{Bj},Q^2) &=&
	{1\over 2} \sum_{q=u,d,..} e_q^2 \Delta q_q(x_{Bj},Q^2).
\end{eqnarray} 
The quark distribution functions (containing quark and antiquark)
are defined as matrix elements of leading  twist-2 operators:
\begin{eqnarray}
\label{deftw2DF}
\Delta q_q(x,Q^2)= \int {d\kappa \over 2\pi (\tilde{x}S)} 
			\left\langle P,S\left|
		O_q^{\rm tw 2}(0,\kappa) + (\kappa \rightarrow -\kappa)
		 	\right|P,S \right\rangle_{|\mu^2=Q^2}
				e^{i x\kappa (\tilde{x}P)},
\end{eqnarray}
where the renormalization point $\mu$ of the operator is set equal to the
momentum transfer $Q$. The twist-2 operators are defined as 
\begin{eqnarray}
\label{p1.8}
O_q^{\rm tw 2}(\kappa_1,\kappa_2) =
				{\bar\psi}_q(\kappa_1 \tilde x)
					\not\!\tilde{x} \gamma_5 
				\psi_q(\kappa_2 \tilde x),
\end{eqnarray}
where $\tilde{x}$ is a light-like vector. Here and in the following we apply
for simplicity the light-cone gauge, i.e. $\tilde{x}A=0$. A gauge invariant 
definition in a general gauge is obtained by including a path-ordered link 
factor.

The operator product analysis to leading order suggests that the 
structure function $g_2$ can be written in the form of a generalized
Wandzura-Wilczek relation \cite{WanWil77}:
\begin{eqnarray}
\label{defg2}
g_2(x_{Bj},Q^2)=
		-\bar{g}_2(x_{Bj},Q^2) 
		+\int_{x_{Bj}}^1 {dy \over y}\, \bar{g}_2(y,Q^2),
\end{eqnarray}
so that the Burkhardt-Cottingham sum rule \cite{BurCot70}, i.e. 
$\int_0^1 dx\, g_2(x,Q^2)=0$, is obviously fulfilled.
Here $\bar{g}_2(x,Q^2) = g_1(x,Q^2) +  \tilde{g}_2(x,Q^2)$ is decomposed 
in the twist-2 part given by $g_1$ and a remaining, not-power suppressed,
twist-3 part $\tilde{g}_2$ \cite{GeyMueRob96b}.  
The twist-3 contribution
\begin{eqnarray}
\label{LOg2}
\tilde{g}_2(x,Q^2) &=&
	{1\over 2} \sum_{q=u,d,..} e_q^2 \Delta \tilde{q}_q(x,Q^2)
\end{eqnarray} 
can now be formally expressed as
\begin{eqnarray}
\label{deftw3DF}
\Delta \tilde{q}_q(x,Q^2)= 
		{1\over x}\int {d\kappa \over 2\pi (\tilde{x}P)} S^\rho
				\left\langle P,S\left|
				\tilde{O}^{\rm tw3}_\rho(0,\kappa) - 
				(\kappa \rightarrow -\kappa)
				 \right|P,S \right\rangle_{|\mu^2=Q^2} 
			e^{i x\kappa (\tilde{x}P)},
\end{eqnarray}
where $S^\rho$ is the polarization vector and the twist-3 light-ray operator 
is defined as
\begin{eqnarray}
\label{p1.9b}
\tilde{O}^{\rm tw3}_{q\rho}(\kappa_1,\kappa_2) &=& 
     \int_0^{1} du\, \bar{\psi}_q(\kappa_1\tilde{x}) i
     \left[
   		\gamma_\rho \tilde{x}_{\sigma}- \not\! \tilde{x}\, 
		g_{\rho\sigma}
	\right] \gamma^5
     D^{\sigma}(u,\kappa_1\tilde{x},\kappa_2\tilde{x})
      \psi_q(\kappa_2 \tilde{x}),
\\
\mbox{with\ }\hspace{1cm} D^\rho(u,\kappa_1\tilde{x},\kappa_2\tilde{x})&=&
				\partial^\rho_{\kappa_2 \tilde{x}}+ 
 	ig A^\rho([\kappa_1 \bar{u}+\kappa_2 u]\tilde{x}), 
	\quad \bar u = 1 - u.
\end{eqnarray}

Unlike $\Delta q_q(x,Q^2)$ this new twist-3 function $\Delta
\tilde{q}_q(x,Q^2)$ has no simple parton interpretation. In fact, using the
equation of motion $(i\!\not\!\!D -m)\psi=0$ it turns out that the twist-3
operator (for more details see \cite{GeyMueRob96a}) can be decomposed in a
basis of three-particle operators containing also the gluon field and in a
mass-dependent two-particle operator. Introducing an appropriate definition
of a three-particle correlation function the net contribution to
$\tilde{g}_2$ in leading order can be written as
\begin{eqnarray}
\label{reltog2}
x \tilde{g}_2(x,Q^2) = 
		{1\over 2} \sum_{q=u,d,..} e_q^2\left[
		{1\over x}m_q(x,Q^2) + 
		{d\over dx} \int_0^1 du\, u \Delta\tilde{q}_q(x,u,Q^2)
											\right].
\end{eqnarray}  
The three-particle correlation function  
\begin{eqnarray}
\label{def3part} 
\Delta\tilde{q}_q(x,u,Q^2) &=& 
		 \int {d\kappa\over 4\pi}  {S_\rho \over (\tilde{x}P)^2} 
			\langle P,S|
		{Y_q^{\rho}}(-\kappa u, \kappa \bar{u})+(\kappa\to-\kappa)
			|P,S\rangle_{|\mu^2=Q^2} 
			e^{i\kappa x (\tilde{x}P)}
\end{eqnarray}
is even under charge conjugation and depends on the variable $u$, which
gives the relative position of the gluon field on the light cone and on the
variable $x$. For $0\le u \le 1$ the gluon field lies between the two quark
fields. Because of the support property $|x|\leq {\rm Max}(1,|2u-1|)$, the
variable $x$ is then restricted to $|x|\leq 1$ and can be interpreted as an
effective momentum fraction. The introduced ligth-ray operator 
$Y_q^{\rho}(\kappa_1,\kappa_2)={^+\!S_q^\rho}(\kappa_1,0,\kappa_2) + 
				        {^-\!S_q^\rho}(\kappa_2,0,\kappa_1)$  
is the non-local generalization of the so-called Shuryak-Vainshtein operators 
\cite{ShuVai82aShuVai82b}:
\begin{eqnarray}
\label{defSVO}
{^\pm\! S_q^\rho}(\kappa_1,\tau,\kappa_2)&=& ig
    \bar{\psi}_q(\kappa_1\tilde{x})\not\!\tilde{x}
     \left[
      i \tilde{F}^{\alpha\rho}(\tau\tilde{x}) \pm
      \gamma^5 F^{\alpha\rho}(\tau\tilde{x}) 
     \right]\tilde{x}_\alpha
    \psi_q(\kappa_2\tilde{x}),
\end{eqnarray}
where 
$\tilde{F}_{\alpha\beta} = {1\over 2} \epsilon_{\alpha\beta\mu\nu} F^{\mu\nu}$
is the dual field strength tensor. Furthermore, we introduced a two-particle 
distribution function 
\begin{eqnarray}
\label{def2part} 
m_q(x,Q^2) &=& 
		\int {d\kappa\over 2\pi}  {S_\rho \over (\tilde{x}P)^2}
			\langle P,S| 
				{M_q^{\rho}}(0,\kappa)+(\kappa\to -\kappa)
			|P,S\rangle_{|\mu^2=Q^2} 
				e^{\{i\kappa x (\tilde{x}P)\}},
\end{eqnarray}
where the operator 
\begin{eqnarray}
\label{moper}
M_q^\rho(\kappa_1,\kappa_2)= 
        m_q\; \bar{\psi}_q(\kappa_1\tilde{x})
     \sigma^{\alpha\rho}\tilde{x}_\alpha \gamma^5
     (\tilde{x}D)(\kappa_2\tilde{x}) \psi_q(\kappa_2\tilde{x}),
\quad \sigma_{\alpha\beta}={i\over 2} [\gamma_{\alpha},\gamma_{\beta}]
\end{eqnarray}
is proportional to the current mass $m_q$.

Using the definition (\ref{def3part}) the evolution equation for the twist-3 
correlation function can be obtained  in a straightforward manner from the 
renormalization group equation (RGE) of the non-local ligth-ray operators. 
From the non-singlet result given in \cite{BalBra89,GeyMueRob96a} one obtains 
an extended DGLAP equation \cite{GeyMueRob96b}:
\begin{eqnarray}
\label{EvoEqux&u}
Q^2{d\over d Q^2} \Delta\tilde{q}^{\rm NS}\left(y,u,Q^2\right) &=& 
	{\alpha_s\left(Q^2\right)\over 2\pi} 
	\int {dz\over z} \int dv
	\Bigg\{
		\tilde{P}_{qq}^{\rm NS}(z,u,v)\,
		\Delta\tilde{q}^{\rm NS}\left({y\over z},v,Q^2\right) + 
\nonumber\\
	&&\hspace{2.5cm}
		\delta(v-u-\bar{u}z) {z\over |v| }\tilde{P}_{qm}(z) 
		\,m^{\rm NS}\left({y \over v},Q^2\right)
	\Bigg\}
\nonumber\\
Q^2{d\over dQ^2} m^{\rm NS}(y,Q^2) &=& 
	{\alpha_s(Q^2)\over 2\pi} \int {dz\over z} \tilde{P}_{mm}(z) \,
	m^{\rm NS}\left({y\over z},Q^2\right).
\end{eqnarray} 
Here, the integration region is determined by both the support of 
$\Delta\tilde{q}^{\rm NS}$ and by the kernel
\begin{eqnarray}
\label{EvoKer}
&&\tilde{P}^{\rm NS}_{qq}(z,u,v) = 
\nonumber\\ 
&&{2C_F-C_A \over 2}
\Bigg[
	 [\Theta_1(z,u,v) K(z,u,v)]_+  +	
	\Theta_2(z,\bar{u},\bar{v}) L(z,\bar{u},\bar{v})-\Theta_2(z,u,v) M(z,u,v)
\Bigg]+ 
\nonumber\\
&&\hspace{0cm} 
 {C_A \over 2}\left[ 
	\left\{
		 [\Theta_3(z,u,v) N(z,u,v)]_+ +
					{u\rightarrow \bar{u} \choose v\rightarrow \bar{v}}
	\right\}+
	\Theta_3(z,u,v) M(z,u,v)-{7\over 2} \delta(u-v)\delta(1-z)\right],
\nonumber
\\
&&\tilde{P}_{qm}(z,u,v)= 
		C_F (1-z)^2, \qquad
\tilde{P}_{mm}(z) = 
		-2 C_F \left(1+z - \left[{1\over 1-z}\right]_+ \right),
\end{eqnarray}
where 
$[A(z,u,v)]_+=A(z,u,v)-\delta(1-z)\delta(u-v)
				\int dz'\int dv'\, A(z',u,v')$
and 
\begin{eqnarray}
\label{auxfunc}
\Theta_1(z,u,v)&=&
	 \theta(z) \theta(u-z v) \theta(\bar{u}-z\bar{v}),
\qquad
\Theta_2(z,u,v)=
	\theta(-\bar{u}\bar{v}z) \theta(\{1-v z\}\bar{u}) 
	\theta(\{z-u\}\bar{u}),
\nonumber\\
\Theta_3(z,u,v)&=&
	\theta(\bar{u}\bar{z})\theta(\bar{u}\bar{v}z)\theta(\{v z-u\}\bar{u}),
\nonumber\\
K(z,u,v) &=&
		z+
	\left\{
	 	{u^2 \over v (v-u)}\delta(u-z v) +
		{u\rightarrow \bar{u} \choose v\rightarrow \bar{v}}
	\right\},
\quad
L(z,u,v) =
	-{\rm sign}(\bar{u}){\bar{v} u^2 \over \bar{u}^2} \delta(u-z),
\nonumber\\
M(z,u,v) &=&
		{2z(1-z v) \over \bar{u}^3}, 
\quad
N(z,u,v) =
		{{\rm sign}(\bar{u}) \bar{v}\over \bar{u} (v-u)} 
			\left\{
				{\bar{v}\over \bar{u}} \delta(1-z) +
				{u^2\over v} \delta\left(u-z v\right)
			\right\}.
\end{eqnarray}
Note that, due to the evolution, the variable $u$ can no longer be
restricted to the region $0\le u \le1$. The moments with respect to $y$
correspond to matrix elements of operators with definite spin $n$, so that
their evolution equation will be diagonal with respect to $n$. However, for
the remaining mixing problem there is no known analytical solution.
In two limits for $x\to 1$ and (only in the non-singlet channel) also in the
large $N_c$ limit, the evolution equation for $\tilde{g}_2(x,Q^2)$ reduces
to an equation of the DGLAP type \cite{AliBraHil91}.

In the singlet channel, there appears a mixing problem. The $C$-even quark 
operator
\begin{eqnarray}
\label{defsquark}
Y^{\rho}(\kappa_1,\kappa_2)=\sum_{q=u,d,\dots} Y_q^{\rho}(\kappa_1,\kappa_2)=
			\sum_{q=u,d,\dots} {^+\!S_q^\rho}(\kappa_1,0,\kappa_2) + 
			{^-\!S_q^\rho}(\kappa_2,0,\kappa_1)
\end{eqnarray} 
will be mixed with the antisymmetric gluon operator:
\begin{eqnarray}
\label{defglu}
G^\rho(\kappa_1,\kappa_2) = 
		g f^{abc} \tilde{x}^\alpha\tilde{x}^\beta\tilde{x}^\gamma
	F^a_{\alpha\mu}(\kappa_1\tilde{x}) \tilde{F}^{b\,\rho}_{\beta}(0)
			F^{c\,\mu}_{\gamma}(\kappa_2\tilde{x}),
\quad G^\rho(\kappa_2,\kappa_1) = -G^\rho(\kappa_1,\kappa_2).					
\end{eqnarray}

\section{Calculation of twist-3 evolution kernels}
\label{Calculation}

The relevant Feynman diagrams for the calculation of the evolution kernel to
leading order are shown in Fig.\ 1. In addition to the one-particle
irreducible diagrams, the application of the equation of motion can be
represented by reducible diagrams \cite{BukKurLip84b}. In the following we
describe shortly the computational calculation of the twist-3 evolution
kernels in the light-cone gauge. The applied method can be easily modified
for a covariant gauge and can be extended to twist higher than 3.

Because of the light-cone gauge all operator vertices 
\begin{eqnarray}
\label{vertices}
O_v(\kappa_1,\kappa_2) = 
	o_v e^{-i\kappa_1(\tilde{x}p_1)-i\kappa_2(\tilde{x}p_2)} 
\qquad\mbox{with}\quad 
	v=\{^+\!S,^-\!S,G\}
\end{eqnarray}
have only a pure exponential $\kappa$-dependence. Here $o_v$ includes the 
Lorentz, flavour, and colour structure of the operator. However, the 
gluon propagator contains a spurious pole:
\begin{eqnarray}
\label{gluonpr}
-i{g^{\mu\nu}-\left(k^{\mu} x^{\nu} + k^{\nu} x^{\mu}\right)/\tilde{x}k
 \over k^2 + i\epsilon},
\end{eqnarray}
which will be regularized by the Leibbrandt-Mandelstam prescription
\cite{Lei84Man83}:
\begin{eqnarray}
\label{LeibManpre}
 \frac{1}{k\tilde{x} }= \frac{k \tilde{x}^\star}
   {(k\tilde x)(k\tilde{x}^\star) +i\epsilon}
\end{eqnarray}
with the dual light-like vector $\tilde{x}^\star$, i.e.
$\tilde{x}^2=\tilde{x}^{\star 2}=0$ and $\tilde{x}\tilde{x}^\star \not=0$.

The Dirac algebra can be performed with one of the usual high-energy
programs (such as FORM, FeynCalc or Tracer). The integration over the loop
momentum and the final simplification of the result were done with a
rule-based program written in Mathematica. After cancellation of common
scalar products between numerator and denominator, as well as the
decomposition of $1/\tilde{x}k...1/\tilde{x}(k+p_i)$, each diagram is
expressed in terms of the following dimensional regularized tensor
integrals:
\begin{eqnarray}
\label{integrals}
I^{\mu_1\cdots \mu_j}(a,b,c,\alpha;\kappa)= 
	\int {d^{4-2\epsilon}k\over (2\pi)^{4-2\epsilon}}
		{
		 k^\mu_1\cdots k^\mu_j	
			\over 
		(k^2
)^a ((k+p_1)^2
)^b ((k+p_2)^2
)^c
		} 
		{e^{-i\kappa (\tilde{x}k)}\over (\tilde{x}k)^\alpha}.
\end{eqnarray}
With the help of the +-prescription
\begin{eqnarray}
\label{defpluspre}
x^{-\alpha}_+ f(x) = 
		{f(x)-\sum_{i=0}^{\alpha-1} x^i f^{(i)}(0)\over x^\alpha},
\qquad f^{(i)}(0)={d^i\over dx^i} f(x)_{|x=0},
\end{eqnarray}
the integral $I^{\mu_1\cdots \mu_j}(a,b,c,\alpha;\kappa)$ can be decomposed in
\begin{eqnarray}
\label{<???>}
I^{\mu_1\cdots \mu_j}(a,b,c,\alpha;\kappa) = 
	I_{+}^{\mu_1\cdots \mu_j}(a,b,c,\alpha;\kappa) + 
	\sum_{\beta=0}^{\alpha-1} (-i\kappa)^\beta
			 I^{\mu_1\cdots \mu_j}(a,b,c,\alpha-\beta;\kappa=0),	
\end{eqnarray}
where $I_{+}^{\mu_1\cdots \mu_j}(a,b,c,\alpha;\kappa)$ contains no more
spurious poles. The remaining integrals with $\kappa=0$ give only a trivial
contribution to the evolution kernel. Such integrals are common in
light-cone gauge and can be found in \cite{Lei87}. Since the integrand of
$I_{+}^{\mu_1\cdots \mu_j}(a,b,c,\alpha;\kappa)$ is analytic in
$\tilde{x}k$, and $\tilde{x}^2$ vanishes, the momentum integration results
in 
\begin{eqnarray}
\label{parametInt}
&&I_{+}^{\mu_1\cdots \mu_j}(a,b,c,\alpha;\kappa)=
		{i (-1)^{a+b+c} \over (4\pi)^{2-\epsilon} } 
		{\Gamma(a+b+c-j-2+\epsilon)\over \Gamma(a)\Gamma(b)\Gamma(c)}
		\int_0^1 dy \int_0^{\bar{y}} dz\, (1-y-z)^{a-1} 
\nonumber\\
&&\hspace{0,5cm}\times y^{b-1} z^{c-1}
	   	{1\over 2} {\partial \over \partial b_{\mu_1}}\cdots  
		{1\over 2}{\partial \over \partial b_{\mu_j}}
		\left\{b^2-D(y,z)\right\}^{-(a+b+c-j-2+\epsilon)} 
		(\tilde{x}b)^{-\alpha}_+
		e^{i\kappa (\tilde{x}b)}|_{b^\mu=B^\mu(y,z)},
\nonumber\\
&&\mbox{where\ }
B^\mu(y,z)=y p_1^\mu + z p_2^\mu, \quad 
		D(y,z)= y p_1^2  + z p_2^2
.
\end{eqnarray}

Performing the derivation with respect to $b_{\mu_i}$ and taking into
account only the ultra-violet (UV) divergent part provide the result for the
corresponding Feynman diagram. Unfortunately, the output is cumbersome and
must be simplified. In a first step all exponentials appearing in the result
of a given Feynman diagram will be transformed in a unique form in such a
way that the measure of the parameter integrals remains unchanged, for
instance:
\begin{eqnarray}
\label{tran1}
e^{		-i\left\{[\kappa_1 (1-y) + \kappa_2 y] \tilde{x}p_1 +
		 [\kappa_2 (z+y) + \kappa_1 (1-y-z)] \tilde{x}p_2\right\}
		 } 
\;\;\stackrel{z\to 1-y-z}{\Longrightarrow}\;\;
e^{		-i\left\{[\kappa_1 (1-y) + \kappa_2 y] \tilde{x}p_1 +
		 [\kappa_2 (1-z) + \kappa_1 z] \tilde{x}p_2\right\}
		 }.
\end{eqnarray}
Finally, such rational functions in the external momenta as 
\begin{eqnarray}
\label{simp1}
	\int_0^1 dy\int_0^{\bar{y}} dz\, \delta(1-y-z)  
	{
		\tilde{x}p_1+\tilde{x}p_2 			\over 
		\left[y \tilde{x}p_1-z\tilde{x}p_2\right]_+
	}
		e^{
		-i\left\{[\kappa_1 \bar{y} + \kappa_2 y] \tilde{x}p_1 +
		 [\kappa_2 \bar{z} + \kappa_1 z] \tilde{x}p_2\right\}
		 }
\end{eqnarray}
have to be transformed into momenta-independent ones. Applying partial 
integration, this expression can be written in the desired form   
\begin{eqnarray}
\label{simp2}
	\int_0^1 dy\int_0^{\bar{y}} dz\,  
		\left[
		\delta(z) {1\over y_{+}} - \delta(y) {1\over z_{+}} 
		\right]
		e^{
		-i\left\{[\kappa_1 \bar{y} + \kappa_2 y] \tilde{x}p_1 +
		 [\kappa_2 \bar{z} + \kappa_1 z] \tilde{x}p_2\right\}
		 }.
\end{eqnarray}

Summing up the contribution of the diagrams and taking into account 
the renormalization of the quark and gluon fields, the evolution kernels are
given by the $1\over \epsilon$ pole \cite{Hoo73Col74}:
\begin{eqnarray}
\label{guark-evolution}
&& \mu^2 {d\over d \mu^2} Y(\kappa_1,\kappa_2) =
	{\alpha_s\over 2\pi}\int_0^1 dy \int_0^{\bar{y}} dz 
\Bigg\{
		\left(C_F-{C_A\over 2}\right)
	\Bigg[
			y\delta (z) Y(-{\kappa_1}y,{\kappa_2}-{\kappa_1}y) -
\nonumber\\
&&\hspace{4,4cm}
			2z Y(\kappa_1-\kappa_2\bar{z},-{\kappa_2}y)+
    	 	[K(y,z)]_+ Y({\kappa_1}\bar{y}+{\kappa_2}y,
									{\kappa_2}\bar{z}+{\kappa_1}z)  
	\Bigg]  + 
\nonumber\\
  & &
  	{C_A\over2}
	\Bigg[ 
		\left(
			2\bar{z}+[N(y,z)]_+ - {7\over 2} \delta(\bar{y})\delta(z)
		\right) 
			Y({\kappa_1}-{\kappa_2}z,{\kappa_2}y)+ 
        [N(y,z)]_+Y({\kappa_1}y,{\kappa_2} - {\kappa_1}z) 
	\Bigg] -
\nonumber\\
&& 2 N_f\, y\bar{y}\, \delta(z)\,
			Y(\kappa_1 y+\kappa_2 \bar{y},\kappa_1 y+\kappa_2\bar{y})+
    C_F\,\bar{y}^2\,\delta(z)
   			\left[M({\kappa_1}-{\kappa_2}y)+ M({\kappa_2}y-{\kappa_1})\right]-
\nonumber\\
&&  N_f(\kappa_1-\kappa_2)
	\Big[
		(1-y-z+4y z)G(\kappa_1\bar{y}+\kappa_2 y,\kappa_1 z+\kappa_2\bar{z})+
	(1-y-z)\times \nonumber\\
&&\hspace{1,8cm} \{
	G(-(\kappa_1-\kappa_2)(\bar{y}-z),-\kappa_1\bar{y}-\kappa_2 y)-
	G(-\kappa_1 z-\kappa_2\bar{z} ,(\kappa_1-\kappa_2)(\bar{y}-z))
		 \}
	\Big]
\Bigg\},
\\
\label{gluon-evol}
& & \mu^2 {d\over d \mu^2} G(\kappa_1, \kappa_2)=
	{\alpha_s\over 2\pi} C_A\int_0^1 dy \int_0^{\bar{y}} dz 
\Bigg\{
(1-y-z+3 y z)
	\big[
    	G(\kappa_1\bar{y} + \kappa_2 y, \kappa_2 \bar{z} + \kappa_1 z)+
\nonumber\\
& & 	G(\kappa_1 y-\kappa_2,\kappa_1 \bar{z}-\kappa_2)
	\big] +
(1-y-z)
	\big[
     	G(\kappa_1 (\bar{y}-z),\kappa_2-\kappa_1 y)-
	    G(\kappa_2 - \kappa_1\bar{z}, -\kappa_1 (\bar{y}-z))  
     \big]+	 
\nonumber\\
&& 
\left(
		[Q(y,z)]_+ -{8C_A+N_f\over 9C_A}\delta(y)\delta(z)
\right)
	\big[
		G(\kappa_1 \bar{y}+\kappa_2 y,\kappa_2 \bar{z}+\kappa_1 z)/2 + 
        G(\kappa_1(\bar{y}-z),\kappa_2-\kappa_1 z)
\big]-
\nonumber\\
& & {1\over 16}[2 - \delta(y) \delta(z)]
			[
			Y(\kappa_1 y-\kappa_2,\kappa_1 \bar{z}-\kappa_2)-
			Y(\kappa_1 \bar{z}-\kappa_2,\kappa_1 y-\kappa_2)
			]/\kappa_1 -
\nonumber\\
& &	{1\over 16}[2-12 y z \delta(1-y-z) + \delta(y)\delta(z)]
	\big[
		2 Y(\kappa_1\bar{y}+\kappa_2 y,\kappa_2\bar{z}+\kappa_1z)/
												(\kappa_1-\kappa_2)-
\nonumber\\
& &		\{
			Y(\kappa_1 y-\kappa_2,\kappa_1 \bar{z}-\kappa_2)+
			Y(\kappa_1 \bar{z}-\kappa_2,\kappa_1 y-\kappa_2)
		\}/\kappa_1
	\big] - (\kappa_1 \leftrightarrow \kappa_2)
\Bigg\},
\\
\label{mass-evolv}
& &
 \mu^2 { d\over d\mu^2} M^\rho({\kappa}) = 
   {\alpha_s\over 2\pi} 2 C_F
 \int_0^1 dy {1\over (1-y)_+} \left[ y^2  M^\rho({\kappa}y)\right],
\end{eqnarray}
where we used the following $+$-prescriptions:
\begin{eqnarray}
\label{defplus}
{[A(y,z)]_+} &=&
A(y,z) - \delta(y)\delta(z) \int_0^1 dy' \int_0^{\bar{y'}} dz' A(y',z'),
\quad \mbox{for\ } A=\{K,Q\},
\nonumber\\
{[N(y,z)]_+} &=&
N(y,z)-\delta(\bar{y})\delta(z)\int_0^1 dy'\int_0^{\bar{y'}}dz' N(y',z'),
\quad
N(y,z) =	\delta (\bar{y} - z)\,{y^2\over \bar{y}} +
		\delta (z)\,{y\over \bar{y}},
\nonumber\\
K(y,z) &=& 
	1 + \delta (z)\,{\bar{y}\over y}  + \delta (y)\,{\bar{z}\over z},
\quad
Q(y,z)= 
	{1\over 2}\left(
		\delta(z) {\bar{y}^2\over y} +\delta(y) {\bar{z}^2 \over z}
			\right).
\nonumber
\end{eqnarray}
Taking into account the different operator definitions, our massless result
coincides (up to a missing factor $N_f$ in the quark-gluon kernel and an
obvious misprint in the gluon-quark sector) with the evolution kernels
calculated by Balitzky and Braun \cite{BalBra89}, restricted to the forward
case. The Taylor expansion with respect to $\kappa_1,\kappa_2$ provides the
local operators and their anomalous dimension matrix. If we set the missing
upper sum limit in Eq.\ (55) of \cite{BukKurLip84b} to $n$, we agree (up to
a constant normalization factor in the definition of the gluon operator) for
the massless case with the general analytical expression for the anomalous
dimension matrix given by Bukhvostov, Kuraev, and Lipatov. Our mass
dependent terms coincide with their non-singlet result.

\section*{Acknowledgement}

It is a pleasure for me to thank B. Geyer, D. Robaschik, and J. Bl\"umlein
for stimulating discussions. I also wish to thank A.V. Belitzky, V.M. Braun,
and G. Hiller. 
This work was financially supported by the Deutsche Forschungsgemeinschaft 
(DFG).

\newpage


\begin{references}

\bibitem{SMC94E143}
D.~Adams et~al. (SMC),
\newblock {\em Phys. Lett.}, 336B:125, 1994;
K.~Abe et~al. (E143),
\newblock {\em Phys. Rev. Lett.}, 76:587, 1996.

\bibitem{BukKurLip83bBukKurLip84a}
A.P. Bukhvostov, E.A. Kuraev, and L.N. Lipatov,
\newblock {\em JETP Lett.}, 37:482, 1983;
\newblock {\em Sov. J. Nucl. Phys.}, 39:121, 1984.

\bibitem{BukKurLip84b}
A.P. Bukhvostov, E.A. Kuraev, and L.N. Lipatov,
\newblock {\em Sov. Phys. JETP}, 60:22, 1984.

\bibitem{Rat86}
P.G. Ratcliffe,
\newblock {\em Nucl. Phys.}, B264:493, 1986.

\bibitem{BalBra89}
I.I. Balitsky and V.M. Braun,
\newblock {\em Nucl. Phys.}, B311:541, 1989.

\bibitem{JiCho90}
X. Ji and C. Chou,
\newblock {\em Phys. Rev.}, D42:3637, 1990.

\bibitem{KodYasTanUem96}
J.~Kodaira, Y.~Yasui, K.~Tanaka, and T.~Uematsu,
\newblock  {\em Phys. Lett.}, B 387:855, 1996.

\bibitem{GeyMueRob96a}
B.~Geyer, D.~M{\"u}ller, and D.~Robaschik,
\newblock ``Evolution kernels of twist-3 light-ray operators in polarized deep
  inelastic scattering'',
\newblock hep-ph 9606320, to appear in Proceedings of the Workshop on QCD
  and QED in Higher Order, Rheinsberg, 1996.

\bibitem{KodYasUem95}
J. Kodaira, Y. Yasui, and T. Uematsu,
\newblock {\em Phys. Lett.}, B 344:348, 1995.

\bibitem{JogLee76}
S.~Joglekar and B.W. Lee,
\newblock {\em Ann. Phys. (NY)}, 97:160, 1976.

\bibitem{AniZav78}
S.~A. Anikin and O.~I. Zavialov,
\newblock {\em Ann. Phys. (NY)}, 116:135, 1978.

\bibitem{BorGeyHorRob85}
M.~Bordag, B.~Geyer, J.~Ho\v{r}ej\v{s}i, and D.~Robaschik,
\newblock {\em Z.\ Phys.}, C26:591, 1985.

\bibitem{BraGeyRob87}
Th. Braunschweig, B.~Geyer, and D.~Robaschik,
\newblock {\em Ann. Physik}, 44:403, 1987.

\bibitem{BukFro87}
A.P. Bukhvostov and G.V. Frolov,
\newblock {\em Sov. J. Nucl. Phys.}, 45:704, 1987.

\bibitem{GeyDitHorMueRob94}
B.~Geyer, F.~M. Dittes, J.~Ho\v{r}ej\v{s}i, D.~M{\"u}ller, and D.~Robaschik,
\newblock {\em Fortschr. Phys.}, 42:101, 1994.

\bibitem{WanWil77}
W.~Wandzura and F.~Wilczek,
\newblock {\em Phys. Lett.}, 72B:195, 1977.

\bibitem{BurCot70}
H.~Burkhardt and W.N. Cottingham,
\newblock {\em Ann. Phys. (NY)}, 56:453, 1970.

\bibitem{GeyMueRob96b}
B.~Geyer, D.~M{\"u}ller, and D.~Robaschik,
\newblock ``The evolution of the nonsinglet twist-3 parton distribution 
function'',  
\newblock DESY 96-239, hep-ph 9611452, to appear in Proceedings of the 
Third Meeting on the Prospects of Nucleon-Nucleon Spin Physics at HERA, 
JINR, Dubna, 1996.

\bibitem{ShuVai82aShuVai82b}
E.V. Shuryak and A.I. Vainshtein,
\newblock {\em Nucl. Phys.}, B199:451 and
\newblock B201:141, 1982.

\bibitem{AliBraHil91}
A.~Ali, V.M. Braun, and G.~Hiller,
\newblock {\em Phys. Lett.}, B266:117, 1991.

\bibitem{Lei84Man83}
G.~Leibbrandt,
\newblock {\em Phys. Rev.}, D29:1699, 1984;
S.~Mandelstam,
\newblock {\em Nucl. Phys.}, B213:149, 1983.

\bibitem{Lei87}
G.~Leibbrandt,
\newblock {\em Rev. Mod. Phys.}, 59:1067, 1987.

\bibitem{Hoo73Col74}
G.~'t~Hooft,
\newblock {\em Nucl. Phys.}, B61:445, 1973;
J.C. Collins,
\newblock {\em Nucl. Phys.}, B80:341, 1974.

\end{references}
\end{document}